\documentclass{mem}
\usepackage{natbib}\usepackage{txfonts}\usepackage{balance}\usepackage{psfig}
\usepackage{graphicx}
\usepackage[a4paper]{hyperref}
\idline{75}{282}

\begin{document}
\def\teff{$T\rm_{eff }$}
\def\kms{$\mathrm {km s}^{-1}$}
\def\etal{{\it et~al.~}}
\def\eg{{\it e.g.~}}
\title{Accretion modes and jet production in black hole X-ray binaries
}

   \subtitle{}

\author{
E. \,Gallo\inst{1} 
\and R.\ P.\, Fender\inst{1,2}
          }
  \offprints{E. Gallo}

\institute{
Sterrenkundig Instituut `Anton Pannekoek', Universiteit van Amsterdam,
Kruislaan 403, 1098 SJ Amsterdam, the Netherlands;~\email{egallo@science.uva.nl}
\and
School of Physics \& Astronomy, University of Southampton,
Highfield, Southampton SO17 1BJ, United Kingdom;~\email{rpf@phys.soton.ac.uk}
} 
 
\authorrunning{Gallo \& Fender}

\titlerunning{accretion and jets in BHXBs}

\abstract{
We review our current understanding of the radio properties of black hole
X-ray binaries in connection with the X-ray spectral states, and discuss them
in the framework of the recently proposed unified model for the jet-accretion
coupling in these systems.
}
\maketitle{}

\section{X-ray states of black hole binaries}

The spectral-energy distribution (SED) of the gravitational power released as
radiation when gas accretes on to a black hole is far
from unique. Different accretion modes are possible, and the same
initial conditions at the outer boundaries may admit more than one solution for
the accretion flow at the inner boundary, with often 
different radiative properties. The main goal of accretion flows theory is to
understand and distinguish all the possible different modes of accretion, and
classify the observed sources in terms of such modes.  The energy spectra of
black hole X-ray binaries (BHXBs) at energies greater than 10 keV are roughly
described by a power law, which may or may not have a detectable high energy
cutoff. The slope of this power-law is characterized by the photon index,
$\Gamma$, where the photon number flux per unit energy (${\rm
photons~cm^{-2}~s^{-1}~keV^{-1}}$) is $F_{\rm N}(E)\propto E^{-\Gamma}$, where
$E$ is the photon energy. 
The power spectra of the X-ray light curves provide
an estimate of the variance as a function of Fourier frequency $\nu$
(typically in the range mHz-kHz) in terms of the power density $P_{\nu}(\nu)$
(see van der Klis 2005 and references therein). Broad -- and hence aperiodic
-- structures in the power spectra are referred to as `noise' while narrow
features are called `quasi-periodic oscillations' (QPOs).

Different X-ray states are distinguished based upon the properties of the
power (strength of the noise, presence or absence of peculiar QPOs) and energy
spectra (broadband luminosity, relative contribution to the X-ray luminosity
of the hard power-law component with respect to a 'soft', quasi-thermal
component which peaks around 1 keV). At luminosities close to the Eddington
one, BHXBs are often in the {`very high state'} (VHS), where both of the two
components contribute substantially to the SED.  At slightly-lower
luminosities, the quasi-thermal component dominates and the power-law is
usually steeper ($\Gamma > 2$) and extended to the $\gamma$-ray band. This
state is traditionally termed {`high/soft'} (HS).  At even lower luminosities,
typically below a few per cent of Eddington,
the 
spectra are completely dominated by a hard power-law component (with
$\Gamma\simeq 1.7$), with the quasi-thermal component extremely weak or even
absent: these are the so-called {`low/hard'} states (LS). Sometimes, at
luminosities intermediate between those of the soft and the hard states, an
{`intermediate state'} (IS) is observed, with properties similar to those of
the 
very high state. Below a few $10^{-5}$ Eddington, a {`quiescent state'} is
identified, with properties similar to the low/hard state.  
In terms of power
spectra, the low/hard, intermediate and very high state are generally
characterized by the presence of strong band-limited noise (i.e., that steepens
towards higher frequencies) and a hard power-law component in the power
spectra, whereas the high/soft state is characterized by these features being
very weak or even absent.  It is often the case that the same source,
either persistent or transient, undergoes a transition between spectral
states, and therefore between accretion modes.

It is generally believed that the main parameter driving the transition
between states is the instantaneous accretion rate $\dot{m}$, even though a
`two-dimensional behaviour' has recently emerged, suggesting that a second
parameter may play a role (Homan \etal 2001).

There are a number of extensive reviews describing in detail the
properties of X-ray states of BHXBs; we refer to the reader to: Esin (1997);
Done (2001); Homan \etal 2001; McClintock \& Remillard 
(2005); Homan \& Belloni (2005). In particular, McClintock \& Remillard (2005)
have introduced a 
new classification that is partly based on the `old' scheme described above,
but no longer 
uses luminosity as a selection criterion. They still recognize a quiescent
state, hard state, soft state (renamed `thermal dominant') and a very high
state (renamed  `steep power law' state) but drop the intermediate one
as a {\it bona fide} state. 

\section{Accretion modes}

The majority of spectral studies of BHXBs in the X-ray band 
suggest that the power-law continua of these sources are produced by 
thermal Comptonization (Shapiro \etal 1976; Sunyaev \&  
Titarchuk 1980) 
in a hot, rarefied `corona' of electron and positrons, which probably resides
where 
most of the accretion energy is released, namely in the inner part of the
flow. Furthermore, there is evidence that this hot, Comptonizing medium
strongly interacts with the colder thermal component: such an interaction is
not only required to explain the ubiquitous reflection features in the X-ray
spectra (Lightman \& White 1988; Matt, Perola \& Piro
1991; Fabian \etal 2000), but could also provide the
feedback mechanism that forces the observed values of coronal temperature and
optical depth to lie in very narrow range for all the different observed
sources (Haardt \& Maraschi 1991).  

From the theoretical point of view, the soft quasi-thermal component is 
thought to be the clear signature of a geometrically thin, optically thick
accretion disc (Shakura 
\& Sunyaev 1973; Pringle 1981). 
Thus, the observed hard-X-ray power laws represent a universal signature of 
a {\it physical process} more than     
of specific accretion dynamics. This is why, if there is little doubt that the
standard thin accretion disc model accounts for the basic
physical properties of black holes in their soft states, the accretion
mode responsible for the low--luminosity hard/quiescent states is still a
matter of debate. 
Radiatively inefficient accretion can take place at low accretion
rates if the density of the accreting gas is low enough to inhibit the energy
coupling between protons and electrons. Under such conditions the flow
remains hot, assumes a puffed-up geometry and radiates very
inefficiently.

Since their rediscovery in recent years (Narayan \& Yi 1994, 1995; Narayan,
Mahadevan \& Quartaet 1998), 
radiatively inefficient, advection-dominated accretion flows
(Ichimaru 1977; Rees \etal 1982) have been regarded as natural
solutions.   
The key feature of an `ADAF' is that the radiative 
efficiency of the accreting gas is low, so that the bulk of the viscously
dissipated energy is stored in the gas as thermal energy (or entropy). ADAF
solutions only exist below a critical accretion rate, $\dot{m}<10^{-2}-10^{-1}
\dot{m}_{\rm Edd}$.
The optically thin gas in an ADAF radiates with a spectrum that is
very different from the blackbody-like spectrum of a thin disc;
more importantly, the luminosity of an ADAF has a steep
dependence on the accretion rate.
The efficiency with which thermal energy is transferred from ions to
electrons (to be subsequently radiated) is proportional to $\dot{m}$, hence
$L\propto \dot{m}^2$.  The quadratic scaling arises because the gas is in
the form of a two-temperature plasma, with the   
ions being much hotter than the electrons. In contrast, the luminosity of a 
Shakura-Sunyaev disc varies
as $L\propto \dot{m}$. 
The key difference
is that whereas in a thin disc a large fraction of the released energy is
radiated, in an ADAF nearly all the energy remains locked up in the gas as
thermal energy and may be advected into the central object. 
When tested against the best data for hard state BHXBs, though,
as in the case of XTE J1118+480 (Esin et al. 2001) or Cygnus X-1 (Esin et
al. 1998), ADAF models alone cannot work. A transition between an inner ADAF
and an outer Shakura-Sunyaev disc is needed, as can also be inferred from
studies of X-ray reflection components (Esin, McClintock \& Narayan 1997; Done
2001).  

There are also concerns with this solution, the main one being that
the accreting gas is generically unbound and can escape freely to infinity. The
reason is that the gas is likely to be supplied with
sufficient angular momentum to orbit the hole and its inflow is controlled by
the rate at which angular momentum is transported outward. This angular
momentum transport is necessarily
associated with a transport of energy. If one attempts to conserve mass, angular
momentum and energy in the flow, it is found that the Bernoulli function -- the
energy that the gas would have if it were allowed to expand adiabatically to
infinity -- is twice the local kinetic energy.

Blandford \& Begelman (1999) have proposed an alternative solution called 
adiabatic inflow outflow solution (ADIOS). Here the key notion is that the
excess energy and angular momentum is lost to a wind at all radii. 
This mass loss makes the accretion rate on to the
black hole much smaller than the rate at which mass is supplied at the outer
radius.  
In this model the radial energy
transport drives an outflow that carries away mass, angular momentum and
energy, allowing the disc to remain bound to the hole.
The final accretion rate into the hole may be only a tiny
fraction (in extreme cases $10^{-5}$) of the mass supply at large radius. 
This leads to a much smaller luminosity than
would be observed from a `conservative' flow. This is important from an
observational perspective, because different assumptions concerning the extent
and nature of the outflow affect the derived densities and temperatures of
the emitting regions, and can lead to very different conclusions based on
phenomenological fits to multi-wavelength data.

Another possible scenario for low-luminosity black holes 
is that proposed by Merloni \& Fabian (2002), where strong, unbound, magnetic
coronae are powered by thin discs at low  
accretion rates. These coronal-outflow-dominated
solutions are both thermally and viscously stable, as in general are all
standard Shakura-Sunyaev accretion disc solution in the gas pressure dominated
regime. However, rapid and dramatic variability in the observed high-energy flux is
expected, as X-rays are produced by coronal structures that are
the eventual outcome of the turbulent magnetic field generation inside the
disc. The geometry of these structures (open {\it vs} closed field lines, for
example) plays a very important role and may be such that, at times, parts of
the corona become temporarily radiatively efficient. 

It remains to be seen which, if any, of these models comes
closest to reproducing the observational characteristics of accretion
on to black holes at different accretion rates. 
Nevertheless numerical
simulations of radiatively inefficient accretion flows also seem to
form outflows, either collimated or not (Hawley \& Balbus 2002).  
\begin{figure*}[t!]
\resizebox{\hsize}{!}{\includegraphics[clip=true]{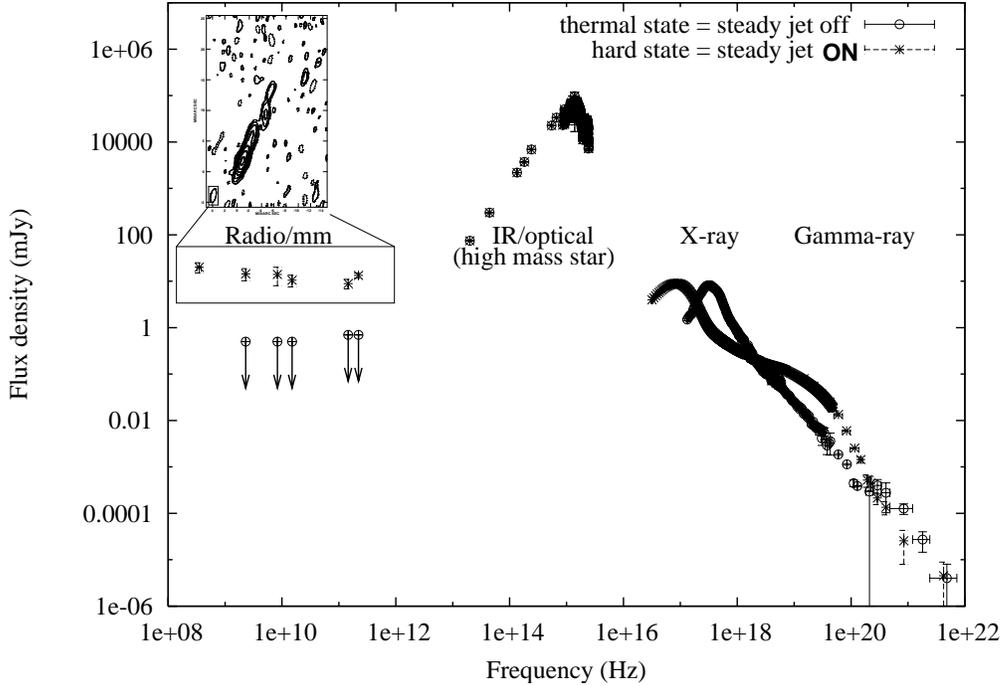}}
\caption{\footnotesize
Spectral energy distribution of the prototypical 10
$M_{\odot}$ BH in the high mass X-ray binary Cygnus X-1, over different accretion regimes. 
When the hard (above $10^{18}$ Hz) X-ray
spectrum is dominated by a hard power-law component, the system is
persistently detected in the radio band. The radio-mm spectrum is flat, due to
a partially self-absorbed steady jet resolved on milliarcsec-scales (inset VLBA map
from \citealt{stirling}). Above a
critical X-ray luminosity of a few per cent Eddington, the disc contribution
becomes dominant (in units of 
$\nu F_{\nu}$), while the hard X-ray power law softens. In this `high/soft', or `thermal
dominant' state the radio emission is quenched by a factor of at least 30 with
respect to the hard state.
}
\label{sed}
\end{figure*}
%
\section{The fifth element}
There also exists a fifth element: the region extending from around the black hole and the
inner edge of the accretion disc and extending out along the symmetry axis to
form a {\it jet}: a narrow stream of energy and particle flowing out of the system 
with relativistic velocities. 
Although there is a general consensus that the formation and initial
collimation of jets requires magnetic fields, we still lack a comprehensive
theory that might account for the process of jet formation, acceleration and collimation.
We do not understand why certain sub-classes of object produce
powerful jets whilst others do not; the composition 
of jets is also a matter of debate: relativistic electrons and magnetic
field must be present, but it is unclear whether the positively-charged
particles are protons or positrons (see e.g. Hughes (1991) for a thorough
review of astrophysical jets).

The advantage of studying relativistic jets powered by
stellar mass objects is simply given by their rapid variability: as the
physical timescales associated with the jet formation are thought to be set by
the accretor's size, and hence mass, then by observing BHXBs on timescales of days to
decades we are probing the time-variable jet:accretion coupling on timescales
of tens of thousands to millions of years or more for supermassive black holes
at the centres of active galactic nuclei.

\section{Radio emission from black hole X-ray binaries} 

Historically, the key observational aspect of X-ray binary jets lies in their
synchrotron radio emission (Hjellming \& Han 1995; Mirabel \& Rodr\'\i guez 1999; Fender
2005).    
The {\it synchrotron} nature of the radio emission from X-ray binaries in
general is inferred by the high brightness temperatures, high degree
of polarization and non-thermal spectra. The
{\it outflow} nature of this relativistic (as it emits synchrotron radiation)
plasma is inferred by brightness temperature arguments, leading to minimum
linear sizes for the emitting region that often exceed the typical orbital
separations, making it unconfinable by any known component of the
binary. 

Different jet properties are associated with different X-ray spectral states
of BHXBs. 
This is illustrated schematically in Figure~\ref{sed}, which
shows the spectral energy distribution, from radio to $\gamma$-ray
wavelengths, of the (prototypical) stellar mass black hole in Cygnus X-1 over
different accretion regimes.

\subsection{Steady jets}

BHXBs in hard states display persistent radio emission with flat radio-mm spectrum. 
Since we are in presence of a relativistic outflow, which is inevitably subject to expansion
losses, the persistence of the emission implies the presence of a continuously
replenished  
relativistic plasma.  The flat spectral indexes can only be produced by
inhomogeneous sources, with a 
range of optical depths and apparent surface brightness, and therefore are
generally interpreted in terms of synchrotron emission from a partially
self-absorbed, steady jet which becomes progressively more transparent al
lower frequencies as the particles travel away from the launching site
 (Blandford \& K\"onigl 1979; Hjellming \& Johnston
1988; Falcke \& Biermann 1996). 
We shall refer to them as {\it steady jets}.
Confirmations of the {\it collimated} nature of these hard state outflows come
from Very Long Based Array (VLBA)  
observations of Cyg X-1 (Stirling \etal 
2001) and GRS 1915+105 (Dhawan
\etal 2000; Fuchs \etal 2003), showing milliarcsec-scale (tens of A.U.)
collimated jets. 
The presence of a steady jet can
also be inferred by its long-term action on the local interstellar medium, as
in the case of the hard state BHXBs 1E1740.7$-$2942 and GRS~1758$-$258, both
associated with arcmin-scale radio lobes (Mirabel
\etal 1992; Mart\'\i~\etal 2002).  Further indications for the existence of
collimated outflows in the hard state of BHXBs come from the stability in the
orientation of the electric vector in the radio polarisation maps of GX~339$-$4 over a
two year period (Corbel \etal 2000). This constant position angle, being the
same as the sky position angle of the large-scale, optically thin radio jet
powered by GX 339$-$4 after its 2002 outburst (Gallo \etal 2004), clearly
indicates a favoured ejection axis in the system.

Some authors propose a jet interpretation (rather
than the standard Comptonizing corona) for the X-ray power-law which dominates the
spectrum of BHXBs in the hard/quiescent state (Markoff,
Falcke \& Fender 2001; Markoff \etal 2003). In this model, depending on the location of the 
frequency above which the jet synchrotron emission becomes optically thin to
self-absorption and
the distribution of the emitting particles, a significant fraction -- if not the
whole -- of the hard X-ray photons would be produced in the inner regions of the
steady jet, by means of optically thin synchrotron and synchrotron
self-Compton emission. \\

No core radio emission is detected while in soft (thermal dominant) state: the radio
fluxes are {\it quenched} by a factor up to about 50 with respect to the hard
X-ray state (Fender \etal 1999; Corbel \etal 2001), probably corresponding to the
physical disappearance of the steady jet. 

\begin{figure*}[t!]
\resizebox{\hsize}{!}{\includegraphics[clip=true]{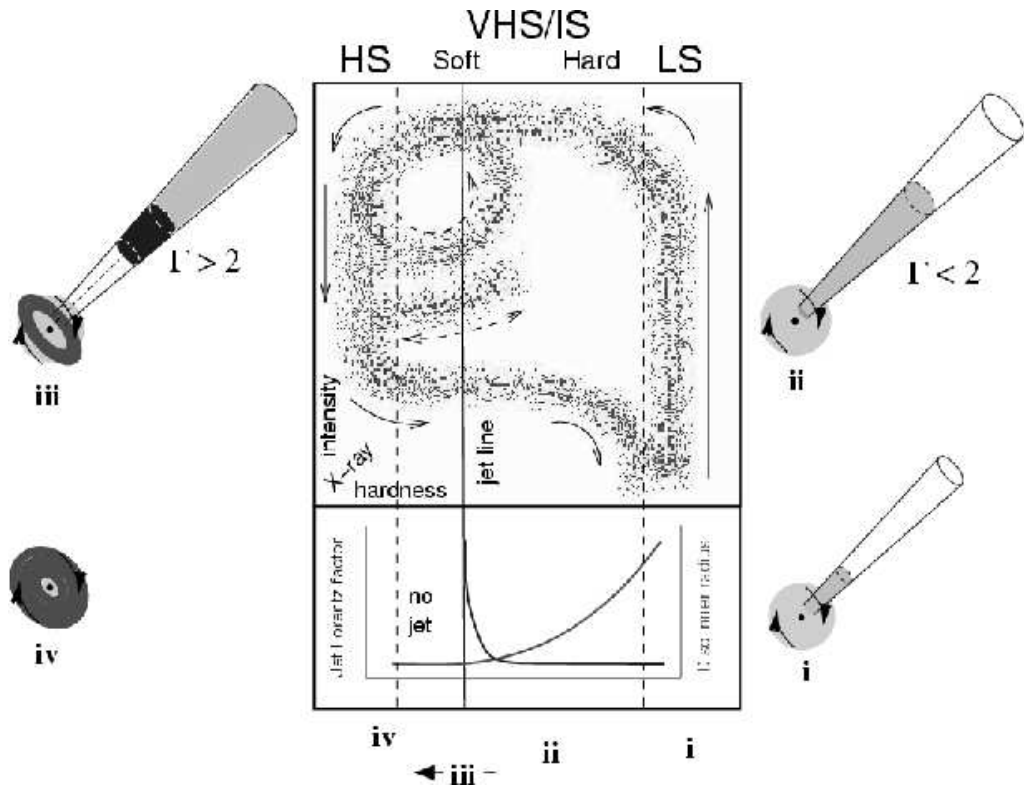}}
\caption{\footnotesize 
A schematic of our simplified model for the jet-disc coupling
  in black hole binaries. The central box panel represents an X-ray
  hardness-intensity diagram (HID); `HS' indicates the `high/soft
  state', `VHS/IS' indicates the `very high/intermediate state' and
  `LS' the `low/hard state'. In this diagram, X-ray hardness increases
  to the right and intensity upwards. The lower panel indicates the
  variation of the bulk Lorentz factor of the outflow with hardness --
  in the LS and hard-VHS/IS the jet is steady with an almost constant
  bulk Lorentz factor $\Gamma < 2$, progressing from state {\bf i} to
  state {\bf ii} as the luminosity increases. At some point -- usually
  corresponding to the peak of the VHS/IS -- $\Gamma$ increases
  rapidly producing an internal shock in the outflow ({\bf iii})
  followed in general by cessation of jet production in a
  disc-dominated HS ({\bf iv}). At this stage fading optically thin
  radio emission is only associated with a jet/shock which is now
  physically decoupled from the central engine.  As a result the solid
  arrows indicate the track of a simple X-ray transient outburst with
  a single optically thin jet production episode. The dashed loop and
  dotted track indicate the paths that GRS 1915+105 and some other
  transients take in repeatedly hardening and then crossing zone {\bf
  iii} -- the `jet line' -- from left to right, producing further optically thin radio
  outbursts. 
}
\label{model}
\end{figure*}

\subsection{Transient jets}

VLA observations of apparent superluminal motions from 
GRS 1915+105, back in 1994, demonstrated unequivocally that
BHXBs could produce highly relativistic jets (Mirabel \&
Rodr\'\i guez 1994). These kind of events have proved to be rather common among BHXBs.
X-ray state transitions appear to be associated with arcsec-scale (thousands
of A.U.) synchrotron-emitting 
plasmons moving away from the  
binary core with highly relativistic velocities (Mirabel \& Rodr\'\i guez
1994, 1999; Fender \etal 1999). Unlike milliarcsec-scale steady jets, such
discrete ejection events display optically thin synchrotron spectra above some frequency,
from which the underlying electron population can be derived. 
The monotonic flux decay observed after a few days in these transient radio
ejections seems to be primarily due to adiabatic
expansion losses, as the decay rate is the same at
all frequencies. Significant loss of energy through the synchrotron
emission process itself, or via inverse Compton scattering, would result in
a more rapid decay at higher frequencies. The
fact that adiabatic losses dominate indicates that the
synchrotron radiation observed from such events is only a small
fraction of the total energy originally input.
We shall refer to them as {\it transient jets}.

\section{Towards a unified model for BHXB jets}

Based upon a collection of quasi-simultaneous radio/X-ray observations of
BHXBs undergoing X-ray state transitions, Fender, Belloni \& Gallo (2004) 
have attempted to construct a unified, semi-quantitative, model for
the disc-jet coupling in BHXBs. 
The model is summarized in Figure~\ref{model}, which we
describe in detail below. The diagram consists of a schematic X-ray
hardness-intensity diagram (HID) above a schematic indicating the bulk
Lorentz factor of the jet and inner accretion disc radius as a
function of X-ray hardness. The four sketches around the outside of
the schematics indicate our suggestions as to the state of the source
at the various phases {\bf i}--{\bf iv}. The path of a typical X-ray
transient is as indicated by the solid arrows.
\begin{itemize}
\item Phase {\bf i}: Sources are in the low-luminosity low/hard X-ray state (LS), producing a
  steady jet. This phase probably extends down
  to the quiescent state.
\item Phase {\bf ii}: The motion in the HID, for a typical outburst,
  has been nearly vertical. There is a peak in the hard state  after which the
  motion in the HID becomes more horizontal (to the left) and the
  source moves into the `hard' (portion of the) very-high/intermediate state (VHS/IS). Despite
this softening of the  
  X-ray spectrum the steady jet persists, with a very similar
  coupling, quantitatively, to that seen in the hard state. 
\item 
Phase {\bf iii}: The source approaches the `jet line' (the
  solid vertical line in the schematic HID) in the HID between jet-producing
  and jet-free states. As the boundary is approached the jet
  properties change, most notably its velocity. The final, most
  powerful, jet, has the highest Lorentz factor, causing the
  propagation of an internal shock through the slower-moving outflow
  in front of it.
\item
Phase {\bf iv}: The source is in the `soft' (portion of the) VHS/IS or the
  canonical high/soft state (HS), and no jet is produced. For a while following the peak
  of phase iii fading optically thin emission is observed from the
  optically thin shock.
\end{itemize}

Following phase {\bf iv}, most sources drop in intensity in the
canonical HS until a (horizontal) transition back, via the VHS/IS, to
the LS. Some sources will make repeated excursions, such as the loops
and branches indicated with dashed lines in Figure~\ref{model}, back across the jet
line, However, with the exception of GRS 1915+105, the number of such
excursions is generally a few.  When crossing the jet line from
right to left, the jet is re-activated but there is (generally) no
slower-moving jet in front of it for a shock to be formed; only motion
from left to right produces an optically thin flare (this is a
prediction). Subsequently the motion back towards quiescence is almost
vertically downwards in the HID.

The inner disc may subsequently recede, in which case a steady jet is
reformed, but with decreasing velocity and therefore no internal
shocks. If the disc once more moves inwards and reaches the `fast jet'
zone, then once more an internal shock is formed. In fact while jets
are generally considered as `symptoms' of the underlying accretion
flow, we consider it possible that the reverse may be true. For
example, it may be the growth of the steady jet (via e.g. build up
of magnetic field near the innermost stable orbit) which results in the
hardening of the X-ray spectrum, perhaps via pressure it exerts on the
disc to push it back, or simply via Comptonization of the inner disc
as it spreads. \\

In the context of the nature and classification of black hole
states, these states, whether `classical' or as redefined by
McClintock \& Remillard (2005) do not have a one-to-one relation with
the radio properties of the source. It seems that as far as the jet is
concerned, it is on -- albeit with a varying velocity -- if the disc
does not reach `all the way in', which probably means as far as the
innermost stable orbit. The dividing `jet line' may also correspond, at
least approximately, to a singular switch in X-ray timing properties
(Belloni 2004; Homan \& Belloni 2005; Remillard 2005) and may be the single
most important transition in the accretion process. Further study of
the uniqueness of the spectral and variability properties of sources
at this transition point should be undertaken to test and refine the above 
model.


\begin{acknowledgements}
E.G. would like to thank the organizers of this meeting for their warm hospitality 
and for finantial support.
\end{acknowledgements}

\bibliographystyle{aa}

\end{document}